

\input phyzzx.tex
\magnification=\magstep1
\def\ie{{\it i.e.}}

\def\oh{{\textstyle {1\over 2}}}

\def\text{\textstyle}

\def\Im{{\rm Im}}
\def\Re{{\rm Re}}
\def\refmark#1{[#1]}                        

\vsize=23.5 true cm
\hsize=15.4 true cm
\predisplaypenalty=0
\abovedisplayskip=3mm plus6pt minus 4pt
\belowdisplayskip=3mm plus6pt minus 4pt
\abovedisplayshortskip=0mm plus6pt
\belowdisplayshortskip=2mm plus6pt minus 4pt
\normalbaselineskip=12pt
\normalbaselines
\noindent

\def\pdx{{\partial\over{\partial x}}}
\def\U{{\tilde V}}

\vsize=23.5 true cm
\hsize=15.4 true cm

\Pubnum{PUPT- 1405}
\def\refmark#1{[#1]}                        
\REF\itz{C. Itzykson and J.-B. Zuber, {\it Jour. Math. Phys. {\bf 21}} (1980)
 411.}
\REF\kasmig{V. Kazakov and A. Migdal, {\sl Induced QCD at Large N},
Princeton preprint PUPT-1322, May 1992.}
\REF\migd{A. Migdal, {\sl Exact Solution of the Induced Lattice Gauge Theory
at Large N}, Princeton preprint PUPT-1323, June 1992.}
\REF\migphase{A. Migdal, {\sl Phase Transitions in Induced QCD},
Preprint LPTENS-92/22, August 1992.}
\REF\gross{David J. Gross, {\sl Some Remarks About the Induced QCD},
Princeton preprint PUPT-1335, August 1992.}
\REF\makeenko{Yu. Makeenko, {\sl An Exact Solution of Induced Large-N
Lattice Gauge Theory at Strong Coupling}, ITEP preprint YM-2-93, March 1993.}
\REF\bipz{E. Brezin, C.Itzykson, G.Parisi and J.-B. Zuber, {\it Comm. Math.
Phys. {\bf 59}} (1978) 35.}
\REF\dj{S.R. Das and A. Jevicki, {\it Mod. Phys. Lett. {\bf A5}} (1990)
 1639.}
\REF\calogero{F. Calogero, {\it Jour. Math. Phys. {\bf 12}} (1971)
 419.}

\def\tr{\rm Tr}

\titlepage
\title{
\baselineskip=28pt
  {\bf ON THE LARGE N LIMIT OF THE ITZYKSON -- ZUBER INTEGRAL}}
\vskip1.5cm
\author{A. MATYTSIN} \address{Joseph Henry Laboratories \break
Princeton University \break Princeton, New Jersey 08544}
\vskip4cm
\abstract
We study the large N limit of the Itzykson -- Zuber integral and show that
the leading term is given by the exponent of an action functional for the
complex inviscid Burgers (Hopf) equation evaluated on its particular classical
solution; the eigenvalue densities that enter in the IZ integral being
the  imaginary
parts of the
boundary values of this solution. We show how this result can be applied
to ``induced
QCD" with an arbitrary potential $U(x)$. We  find that
for a nonsingular $U(x)$ in one dimension the eigenvalue density
$\rho(x)$ at the saddle point is the solution of  the functional equation
$G_{+}(G_{-}(x))=G_{-}(G_
{+}(x))=x$, where $G_{\pm}(x)
\equiv {1\over{2}}U^{\prime}(x)\pm i\pi \rho(x)$. As an illustration we
present a number of new particular solutions of the $c=1$ matrix model on a
discrete real line.

\vsize=23.5 true cm
\hsize=15.4 true cm
\vfill
\eject
\chapter{ \bf   Introduction}

The analysis of most any multi-matrix model is based on the
Itzykson -- Zuber formula \refmark{\itz}
$$I(\phi, \chi)\equiv\int{\cal D}Ue^{N\tr[\phi U\chi
U^{\dagger}]}={{{\rm det}[e^{N\phi_{i}
\chi_{j}}]}\over{\Delta(\phi)\Delta(\chi)}},
\eqn\izi$$
where $\phi$ and $\chi$ are arbitrary $N \times N$ Hermitian matrices,
$\Delta(\phi)$
denotes the Van der Monde determinant constructed out of the eigenvalues
$\phi_{i}$ of these matrices,
$\Delta(\phi)=\Pi_{i<j}(\phi_{i}-\phi_{j})$,
and ${\cal D}U$ is the Haar measure on
the unitary group $U(N)$.
The way to apply it is to diagonalize the matrices entering the model
and to carry out the integration over the diagonalizing unitary matrices
using \izi.  Then, as $N\rightarrow \infty$,
the  residual integrations (over the eigenvalues
of the matrices) can be treated by the saddle point method. In this limit
the eigenvalues at the saddle point can be described by their
distribution on the real axis: if ${\rm d}i$ is the number of eigenvalues
between
$\phi$ and $\phi +{\rm d}\phi$, then there is a smooth function $\rho(\phi)$
such that ${\rm d}i/N = \rho(\phi) {\rm d}\phi$. If we keep the
distribution $\rho(\phi)$ of the eigenvalues $\phi_{i}$ entering \izi \
(as well as $\sigma(\chi)$ of $\chi_{i}$) fixed, the integral \izi \
posesses  the
$N\rightarrow
\infty$ asymptotic expansion,
$$I(\phi, \chi)\sim \exp\{N^{2}F_{0}(\phi, \chi)+F_{1}(\phi, \chi)+N^{-2}F_{2}
(\phi, \chi)+...\}. \eqn\izas$$
Nevertheless, although there is an exact explicit formula \izi \
 for the Itzykson
-- Zuber integral for any finite N,
the structure of the coefficients $F_{n}$ is not known yet.
For many applications it would be important to understand it.

 An interesting example of such an application is the so-called ``induced QCD"
\refmark{\kasmig}
which is the lattice gauge theory of a scalar matrix field $\Phi(x)$ defined by
the action
$$ S= \sum_{x}  N  \tr \biggl[ -U(\Phi(x)) + \sum_{\mu }  \Phi(x)
U_{\mu}(x)  \Phi(x+\mu a) U^{\dagger}_{\mu}(x)\biggr]. \eqn\km$$
($\mu$ runs over the $2D$ lattice vectors on
a hypercubic lattice of dimension $D$, the field $\Phi({\rm x})$ transforms
according to the adjoint representation of the
gauge group $U(N)$). It describes the high temperature limit
of $D+1$ dimensional QCD and, in addition, can be related to the matrix
models of $c=D$ strings. The Itzykson -- Zuber integral turns out to be
the main computational tool necessary to analyze this theory.

In this paper we claim that
large N limit of the IZ integral is given  by
$$ I(\phi, \chi) \sim {\rm exp}\bigl(N^{2}F_{0}(\phi, \chi)\bigr) , $$
$$\eqalign{F_{0}(\phi, \chi) = S[\rho, \sigma]+
{1\over  2}\biggl\{&\int\rho(x)x^{2}dx +\int \sigma(y) y^{2}dy\biggr\} \cr
-{1\over2}
\biggl\{&\int \rho(x_{1})\rho(x_{2})\ln|x_{1}-x_{2}|dx_{1}dx_{2} \cr  +
&\int \sigma(y_{1}) \sigma(y_{2})\ln|y_{1}-y_{2}|dy_{1}dy_{2}\biggr\} \cr}
\,\,\eqn
\kaldef$$
where $S[\rho, \sigma]$ is the classical action corresponding to the Hopf
equation
$${\partial f\over{\partial t}}+ f \pdx f =0, \eqn\hopf$$
calculated on its solution satisfying the boundary conditions
$$\eqalign{
\Im f(x, t=0)&=\pi \rho(x) \cr \Im f(x, t=1)&=\pi\sigma(x) \cr}\,\,.
\eqn\bondcond
$$
We will show that the action  $S[\rho, \sigma]$ defines the field theory
with the Hamiltonian
$$H[\rho, \Pi]={1\over 2}\int \rho(x)\Biggl\{\Bigl({\partial\Pi(x)\over
{\partial x}}\Bigr)^
{2} -{\pi^{2}\over 3}\rho^{2}(x)\Biggr\}dx  \eqn\ham $$
where $\Pi(x)$ is the canonical momentum conjugate to $\rho(x)$. In this
notation
$$f(x,t)\equiv {\partial \Pi(x,t)\over
{\partial x}}+ i\pi\rho(x,t).\eqn\deffunf $$
Quite remarkably, \ham \ is the
nonunitary version of the Hamiltonian, arising in
the collective field formulism of $c=1$ string theory \refmark{\dj}.

The plan of this paper is as follows. In sections 2 and 3 we establish
the relation between the IZ integral and the Hopf equation, and prove the
boundary conditions \bondcond.
In section 4 we apply our results to analyze ``induced QCD" with an
arbitrary nonsingular potential $U(x)$. We obtain that in
the one-dimensional case the eigenvalue density $\rho(x)$
as a solution of  the master field
equation must satisfy the functional constraint
$G_{+}(G_{-}(x))=G_{-}(G_{+}(x))=x$,
where $G_{\pm}(x)\equiv {1\over{2}}U^{\prime}(x)\pm i \pi \rho(x)$.
Although we were not
able to find an explicit general solution to this equation, we show how
one can generate a number of particular solutions of it.
We use them to analyze the $c=1$ matrix model on a lattice for finite lattice
spacings.

\chapter{\bf Reduction of the Itzykson -- Zuber integral to
the Hopf problem}

One of the ways to approach the problem of the large N expansion is to
write the differential equation for the IZ integral and
specify the corresponding boundary conditions.
So, first we will derive the differential equation and
then identify the relevant boundary conditions.

 It is convenient to consider the
one parameter family of integrals
:
$$I(\phi, \chi|t)\equiv{1\over{ t^{N^{2}\over 2}}}\int{\cal D}
U{\rm e}^{-{N\over 2t}\tr(\phi -U\chi U^
{\dagger})^{2}}, \eqn\izt$$
so that $I(\phi, \chi|t)=\exp\bigl(-{N\over{2t}}\tr(\phi^{2}+\chi^{2})\bigr)
I({\phi\over {\sqrt{t}}}, {\chi\over {\sqrt{t}}})$ and  the
 Itzykson -- Zuber integral per se corresponds to
$t=1$.
Since $I(\phi, \chi|t)$ depends only on the eigenvalues of $\phi$ and $\chi$,
one can show \refmark{\itz} that \izt \ satisfies the partial differential
equation
$$ 2N {\partial I(\phi, \chi|t) \over {\partial t}}={1\over{\Delta(\phi)}}
\sum_{i=1}^{N}{\partial^{2} \over {\partial \phi_{i}^{2}}}\Delta(\phi)
I(\phi, \chi|t),\eqn\heat$$
where $\Delta(\phi)=\Pi_{i<j}(\phi_{i}-\phi_{j})$.

This equation is the linear heat equation for the
quantity ${\tilde I} (\phi, \chi|t)=\Delta(\phi) I(\phi, \chi|t)$ and
can be solved exactly to obtain the Itzykson --
Zuber formula \izi. However, unlike $I(\phi, \chi|t)$,
${\tilde I} (\phi, \chi|t)$ does {\it not} have any
regular large N asymptotics. Therefore we have to study the equation for $I$.
Let us use the ansatz
$$I(\phi, \chi|t)=\exp\bigl(N^{2}W[\rho, \sigma |t]\bigr)$$ so that from
\heat \ follows
$$2{\partial W \over {\partial t}}= {1 \over N}\sum_{i=1}^{N}{\partial^{2}
\over {\partial \phi_{i}^{2}}}W+ N \sum_{i=1}^{N}\biggl(
{\partial W \over {\partial
\phi_{i}}}\biggr)^{2}+2 \sum_{i=1}^{N}V(\phi_{i}){\partial W \over {\partial
\phi_{i}}},\eqn\ww$$
where
$$V(\phi_{i})= {1 \over N}\sum_{j\ne i}{1\over{\phi_{i}-\phi_{j}}}.$$

The equations above are exact.
Now we assume that $W[\rho, \sigma |t]$ has a smooth large N limit, \ie \
it is a smooth functional of the eigenvalue densities $\rho$ and $\sigma$.
This implies that ${\partial W \over {\partial
\phi_{i}}}\sim {\cal O}(1/N)$ and
$${\partial W \over {\partial
\phi_{i}}}= {1\over N}\biggl(
{\partial \over {\partial x}}{\delta W \over {\delta \rho(x)}}\biggr)\biggl|_{
x=
\phi_{i}}. \eqn\lNl$$
Thus we can neglect the first term on the left-hand side of \ww (it is ${\cal
O}(1/N)$ while the two other terms are ${\cal
O}(1)$\footnote*{If we would have been looking for $\tilde W$ such that
${\tilde I}={\rm exp}(N^{2} {\tilde W})$, we would not have been
able to neglect the
first term.})to get
$$2{\partial W \over {\partial t}}=N \sum_{i=1}^{N}\biggl(
{\partial W \over {\partial
\phi_{i}}}\biggr)^{2}+2 \sum_{i=1}^{N}V(\phi_{i}){\partial W \over {\partial
\phi_{i}}},\eqn\www$$
Substituting
$$W=S-{1\over
2N^{2}}\sum_{i\ne j}\ln\bigl|
\phi_{i}-\phi_{j}\bigr|-{1\over
2N^{2}}\sum_{i\ne j}\ln\bigl|
\chi_{i}-\chi_{j}\bigr|  \eqn\relationship
$$
we find
$$2{\partial S \over {\partial t}}=N \sum_{i=1}^{N}\biggl({\partial S \over
{\partial\phi_{i}}}\biggr)^{2} -{1\over N}\sum_{i=1}^{N}V^{2}(\phi_{i}).$$
Finally, we use the identity ${1\over N}\sum_{i=1}^{N}V^{2}(\phi_{i})={1\over
{N^{3}}}\sum_{i\ne j}{1
\over{(\phi_{i}-\phi_{j})^{2}}}$ to obtain

$$2{\partial S \over {\partial t}}=N \sum_{i=1}^{N}\biggl({\partial S \over
{\partial\phi_{i}}}\biggr)^{2} -{1\over
{N^{3}}}\sum_{i\ne j}{1\over{(\phi_{i}-\phi_{j})^{2}}}.\eqn\hamjac$$

It is very useful to rewrite this equation
 in terms of the eigenvalue distributions.
To this end we replace all the sums by the integrals according to the rule
$${1\over N}\sum_{i=1}^{N}\rightarrow \int\rho (x)dx.$$
However, the sum ${1\over
{N^{3}}}\sum_{i\ne j}{1\over{(\phi_{i}-\phi_{j})^{2}}}$
needs to
be treated specially, since the relevant integral $\int\int{\rho(x)dx\rho
(y)dy\over {(x-y)^{2}}}$ diverges. In fact the ${\cal O}(1)$ contribution to
this sum comes from the pairs of the eigenvalues $\phi_{i}$
and $\phi_{j}$ such that $|i-j|\ll N$. Thus $\phi_{i}-\phi_{j}\approx|i-j|/
\bigl(N
\rho(\phi_{i})\bigr)$
and
$${1 \over N}\sum_{j}{1\over{(\phi_{i}-\phi_{j})^{2}}}\sim N \rho^{2}(\phi_{i})
\sum_{j}{1\over{(i-j)^{2}}} = {\pi^{2}\over 3}N \rho^{2}(\phi_{i}).$$
Replacing sums by integrals and using \lNl \  we see that
the continuum version of \hamjac \ is
$$2{\partial S \over{\partial t}}=\int \rho(x)dx \Biggl\{\biggl(
{\partial\over {\partial
x}}{\delta S\over{\delta \rho(x)}}\biggr)^{2} - {\pi^{2}\over 3}\rho^{2}(x)
\Biggr\}.
\eqn\hamjaccont$$
This is a nonlinear differential equation for the functional $S[\rho(x),
\sigma(y) | t]$. It would be extremely hard, if possible at all, to find
its general solution. Fortunately, we have to find only a particular
solution - the one which corresponds to the large N limit of the IZ integral.
To do so we will guess what the desired solution is and then prove the
conjecture.

First note that \hamjaccont \ can be represented in the Hamilton - Jacobi
form
$${ \partial S \over {\partial \tau}}+ H\Bigl({\delta S \over{\delta \rho(x)}},
\rho(x) \Bigr)=0 \eqn\hj $$
where $\tau \equiv -t$,
$$H\bigl[\rho(x), \Pi(x) \bigr]=
{1\over 2}\int \rho(x)\Biggl\{\Bigl({\partial\Pi(x)\over
{\partial x}}\Bigr)^
{2} -{\pi^{2}\over 3}\rho^{2}(x)\Biggr\}dx \eqn\H $$
is the Hamiltonian \footnote*{The discrete version of this Hamiltonian, the one
corresponding to \hamjac, describes the completely integrable Calogero
theory \refmark{\calogero}.}, and $\Pi(x)$ - the canonical momentum
conjugate to
$\rho(x)$. Our conjecture is that $S[\rho(x),
\sigma(y) | \tau]$ is given by the action of the dynamical system defined by
\H \ calculated along the trajectory which connects the two points $\rho(x)$
and $\sigma(y)$ in the configuration space of this system. The trajectory
has to satisfy the equations of motion and take $\rho(x)$ to $\sigma(y)$
within the time interval $\tau$.

By construction the so defined $S[\rho(x),
\sigma(y) | \tau]$
solves  \hj.  One still has to show that it gives the right large N
asymptotics of the IZ integral. This will be done in the next section.
In the meantime we will investigate the properties of this solution in more
detail.

To find $S[\phi, \chi]$ we have to solve the Hamilton equations following from
\H. These are
$$\eqalign{
{\partial \rho(x,t)\over{\partial t}}=&-{\partial\over{\partial x}}
\biggl(\rho(x,t){\partial \Pi(x,t)\over {\partial x}}\biggr) \cr
{\partial \Pi(x,t)\over{\partial t}}=&-{1\over 2}\biggl(
{\partial \Pi(x,t)\over
{\partial x}}\biggr)^{2}+{\pi^{2}\over 2}\rho^{2}(x, t) \cr }\,\,.$$

Introducing $$v(x,t)\equiv{\partial \Pi(x,t)\over
{\partial x}} \eqn\v$$
we get
$$ \eqalign{
&{\partial \rho\over{\partial t}}+{\partial\over{\partial x}}
(\rho v)=0 \cr
&{\partial v \over{\partial t}}+ v \pdx v = {\pi^{2}\over 2} \pdx
\rho^{2}
\cr }\,\,. \eqn\euler$$
These are the Euler equations for a one dimensional liquid obeying
the equation of state $P=-{\pi^{2}\over 2}\rho^{2}(x)$. They can be
easily solved. Indeed, let us introduce a new unknown function
$$f(x,t)\equiv v(x,t)+ i\pi\rho(x,t).$$
Then $f$ satisfies the Hopf equation
$${\partial f\over{\partial t}}+ f \pdx f =0. \eqn\hopfie$$

Now we have to specify the boundary conditions for \euler, \hopfie \ that lead
uniquely to the original problem \izi, \izas. Our conjecture is to identify
the endpoints of the classical solution we are looking for with the
eigenvalue densities of the matrices $\phi$ and $\chi$. Therefore we demand
that
$$\cases{
\rho(x, t=0)= \rho(x) \cr \rho(x, t=1)=\sigma(x) \cr}\,\,
$$
or, equivalently,
$$\cases{
\Im f(x, t=0)=\pi \rho(x) \cr \Im f(x, t=1)=\pi\sigma(x) \cr}\,\,. \eqn\bc
$$

Finally, taking into account that due to \izt \ $F_{0}(\phi, \chi)=
W[\rho, \sigma |t] + {1\over{2N}}\tr(\phi^{2}+\chi^{2})$ and remembering
the definition of $S[\rho, \sigma]$, \relationship , we obtain our final
result \kaldef , \hopf, \bondcond.

Is it possible to find an explicit formula for $F_{0}(\phi, \chi)$ ? To
do this we have to find the trajectory obeying \hopfie, \bc.
The Hopf equation is, in principle, exactly soluble. For a $real$ function $f$
the general solution of \hopf\ can be written down in the parametrized form
$$\eqalign{
x=&R(\alpha)+F(\alpha)t \cr
f(x,t)=&F(\alpha) \cr}\,\,. \eqn\parsol$$
where $R(\alpha)$ and $F(\alpha)$ are the two functions to be determined
from the initial conditions supplementing \hopfie. In fact, because of the
invariance with respect to reparametrizations of $\alpha$, \parsol  \ depends
on only $one$ function, consistent with the fact that the Hopf equation is
of first order.

Our problem is a little more complicated, because $f$ is complex valued. One
can generalize  the parametric solution to describe this situation provided
that the initial condition $f_{0}(x)= f(x, t=0)$ can be analytically continued
to the whole complex plane as a function of $x$. This is indeed true
in many applications.

In this case $F(\alpha)$ and $R(\alpha)$ can be viewed as the analytic
functions
of the complex parameter $\alpha$.
To impose the boundary conditions we introduce another pair of functions,
$$G_{+}(x)=x+f(x, t=0), \, \ G_{-}(x)=x-f(x, t=1)$$
and use \parsol \ to get
$$G_{+}(x)=x+ [F\circ R^{-1}](x) = [(F+R)\circ R^{-1}](x)$$ as well as
$$G_{-}(x)=x- [F\circ (F+R)^{-1}](x)= [R\circ (F+R)^{-1}](x)$$ so that
$$        \cases{ G_{+}(G_{-}(x))=G_{-}(G_{+}(x))=x \cr
                  \Im G_{+}(x)=\pi \rho(x) \cr
                  \Im G_{-}(x)=-\pi\sigma(x) \cr} \eqn\pakost$$
where $\circ$ denotes functional composition.
Given $\rho(x)$ and $\sigma(x)$ we must solve \pakost \  for, say, $G_{+}(x)$,
then the full initial condition for the
Hopf equations \hopfie \ will be given by
$f(x, t=0)= G_{+}(x)-x$. This allows one, in principle, to calculate the
trajectory connecting $\rho(x)$ and $\sigma(x)$ and the action along it,
which, in turn, gives the large N asymptotics of the Itzykson -- Zuber
integral.

 We were not able to obtain an explicit expression for $F_{0}$ in \izas\
as a functional of $\rho$ and $\sigma$.  However, in some cases
\pakost \ already provides enough information.
This is the case in ``induced QCD".
We will return to this question in  section 4.

\chapter{\bf Boundary Conditions}

In this section we will prove our conjecture about
the boundary conditions for the Hopf equation.
We will show that the quantity $S[\rho, \sigma]$ entering \kaldef \
is indeed the classical action along the trajectory of \ham \
 with the endpoints
$$\rho(x,t=0)=\rho(x);\,\, \,\,\ \rho(x,t=1)=\sigma(x). \eqn\bcnd$$
To this end we note that the Itzykson -- Zuber integral
$$I(\phi, \chi)\equiv\int{\cal D}Ue^{N\tr[\phi U\chi
U^{\dagger}]} \eqn\apiz $$
satisfies the set of identities with higher order differential operators
$H^q_N \equiv \Tr\bigl({1\over N}{\partial\over {\partial \phi}}\bigr)^{q}$
$$H^q_NI(\phi,\chi)=
\bigl[\Tr(\chi^{q})\bigr]I(\phi, \chi),\,\, k=1,...,N. \eqn\lapls$$
Since $I(\phi, \chi)$ depends only on the eigenvalues of the matrices $\phi$
and $\chi$, these identities reduce to
$$
{\cal D}_{q}I(\phi, \chi)= \bigl[\Tr(\chi^{q})\bigr]I(\phi, \chi) \eqn\dq$$
where
$${\cal D}_{q}\equiv {1\over{\Delta(\phi)}}\sum _{i=1}^{N}{\partial ^{q}
\over{\partial \phi_{i}^{q}}}\Delta(\phi). \eqn\dqdef$$

To take the large N limit of the differential operator in \dqdef \ we expand
it
explicitly to obtain
$$\eqalign{
{\cal D}_{q}=&\sum_{
{\scriptstyle k_{1},\ldots ,k_{s} \atop {\scriptstyle
 \lambda; n_{1},\ldots,n_{s} \atop{\scriptstyle
n}}}}
{q!(-)^{(k_{1}-1) n_{1} +\ldots+(k_{s}-1)n_{s}}\over{\lambda ! k_{1}^{n_{1}}
\ldots
k_{s}^{n_{s}} n_{1} ! \ldots
n_{s} ! (q-\lambda - n_{1}k_{1}-\ldots-n_{s} k_{s}) !}}
\cr
&
\bigl(\U(\phi_{n})\bigr)^{\lambda}\bigl(Z_{k_{1}}(\phi_{n})\bigr)^{n_{1}}\ldots
\bigl(Z_{k_{s}}(\phi_{n})\bigr)^
{n_{s}}\biggl({\partial \over{\partial \phi_{n}}}\biggr)
^{q-(\lambda + n_{1}k_{1}+\ldots+
n_{s} k_{s})} \cr}\eqn\horrible$$
where $$\eqalign{\U(\phi_{n}) = &\sum_{k\ne n }{1\over {\phi_{n}-\phi_{k}}},\cr
Z_{m}(\phi_{n})=&\sum_{k\ne n }{1\over {(\phi_{n}-\phi_{k})^{m}}} \cr}
$$
In the large N limit $$
\U(\phi_{n})\rightarrow N\,\, {\cal P}
\int{\rho(x)dx\over{\phi_{n}-x}} = {\cal O}(N)$$ where ${\cal P}$ denotes the
principal value prescription,
and, naively, $Z_{m}(\phi_{n})$ would go like ${\cal O}(N)$ as well (since
it includes one summation over $k$). However, it is not
true because of the singularity near $\phi_{n} \sim \phi_{k}$ where
$\phi_{n} -\phi_{k}\approx (n-k)/\bigl(N\rho(\phi_{n})\bigr)$ so that
$$\eqalign{
Z_{m}(\phi_{n})=& N^{m}\rho^{m}(\phi_{n}) \sum_{k\ne n }{1\over{(n-k)^{m}}}
+{\cal O}(N^{m-1}) \cr
&=\cases{2N^{m}\zeta(m)\rho^{m}(\phi_{n}) +{\cal O}(N^{m-1}) ,{\rm if }\,\,
 m\,\, {\rm is\,\, even} \cr
{\cal O}(N^{m-1}) , {\rm if}\,\, m \,\,{\rm \,\,is\,\,
 odd}} \cr} \eqn\zetafun$$
where
$$\zeta(m)=\sum_{k=1}^{\infty}{1\over{ k^{m}}}$$
is the Riemann zeta function.

 When the operators ${\cal D}_{q}$ act on $I(\phi, \chi)=
\exp\bigl(N^{2}F_{0}[\rho, \sigma]\bigr)$, the leading $N\rightarrow\infty$
contribution comes from
$$\biggl({\partial\over{\partial \phi_{n}}}\biggr)^{p} \exp(N^{2} F_{0})=
\Biggl[N^{p}\biggl(N {\partial F_{0} \over{\partial  \phi_{n}}}\biggr)^{p}
+{\cal O}(N^{p-1})\Biggr]
\exp(N^{2} F_{0})$$
and is given by
$$\eqalign{
&{\cal D}_{q}\exp(N^{2} F_{0})=\Biggl\{ N^{q}\sum_{
{\scriptstyle k_{1},\ldots,k_{s} even \atop
{all\, different, all\ge 2
 \atop {
\scriptstyle \lambda; n_{1},\ldots ,n_{s} \atop {\scriptstyle
n}}}}}
{q!(-)^{(k_{1}-1) n_{1} +\ldots
+(k_{s}-1)n_{s}}\over{\lambda ! k_{1}^{n_{1}}...
k_{s}^{n_{s}} n_{1} ! \ldots
n_{s} !}} \cr
&{1\over{ (q-\lambda - n_{1}k_{1}-\ldots-n_{s} k_{s}) !}}
\bigl(\U(\phi_{n})\bigr)^{\lambda} {\prod_{j=1}^{s}\bigl(
2\zeta(k_{j})\rho^{k_{j}}
(\phi_{n})\bigr)^
{n_{j}}} \cr &
\biggl(\pdx {\delta F_{0}\over{ \delta \rho(x)}}\Bigm|_{x=\phi_{n}}\biggr)
^{q-(\lambda + n_{1}k_{1}+\ldots+
n_{s} k_{s})} + {\cal O}(N^{q-1}) \Biggr\}\exp(N^{2} F_{0}) \cr} \eqn\qexpl $$
Performing the summation over $k_{1},\ldots,k_{s};n_{1},\ldots ,n_{s}$ and $s$;
using $${1\over N}\sum_{i=1}^{N}\rightarrow \int\rho (x)dx,$$
as $N\rightarrow \infty$ we obtain
\footnote*{To do the sum it is convenient to introduce an operator
${\cal D}_{\infty}=\sum_{q=0}^{\infty}{(iz)^{q}\over{q!}}{\cal D}_{q}$.
 Then the
summation can be performed using the identity $$\exp\Bigl[-2
\sum_{k=2,4,6,\ldots}{\zeta(k)\over{k}}\bigl(iz\rho(\phi)\bigr)^{k} \Bigr]=
{{\rm sinh}\bigl(\pi z \rho(\phi)\bigr)\over{\pi z \rho(\phi)}}$$
 with the result
$${\cal D}_{\infty}\exp(N^{2} F_{0})= {1\over{\pi z}}\int d\phi \,
 {\rm exp}\Bigl[i z
\bigl( \U(\phi) + {\partial \over{\partial \phi}}
{\delta F_{0}\over{\delta \rho(\phi)}}\bigr)\Bigr] {\rm sinh}
\bigl(\pi z \rho(\phi)\bigr)\exp(N^{2} F_{0}) $$ .  }
$$\eqalign{{\cal D}_{q}\exp\bigl(N^{2} F_{0}[\rho, \sigma]&\bigr) =\cr
{1\over{2\pi i}}{1\over {q+1}}\int dx
\biggl[&\Bigl(\pdx {\delta F_{0}[\rho, \sigma]\over{\delta \rho(x)}}+
\U(x) +i\pi \rho(x) \Bigr)^{q+1} -\cr
\Bigl(\pdx {\delta F_{0}[\rho, \sigma]\over{\delta \rho(x)}}+&
\U(x) -i\pi \rho(x) \Bigr)^{q+1} \biggr]\exp\bigl(N^{2} F_{0}[\rho, \sigma]
\bigr)
 \cr&=\biggl[\int\sigma(y) y^{q}dy\biggr]
\exp\bigl(N^{2} F_{0}[\rho, \sigma]\bigr) \cr }\eqn\const
$$
and thus the large N limit of the identity \lapls\ is given by \const.
If we now substitute ansatz \kaldef \ into \const \  we will get  that the
quantity
$$\Pi(x)= -  {\delta S[\rho, \sigma]\over {\delta \rho(x)}}$$
should satisfy
$$\eqalign{
{1\over{2\pi i}}{1\over {q+1}}\int dx
\biggl[&\Bigl(-\pdx \Pi(x)+
\U(x) +i\pi \rho(x) \Bigr)^{q+1} -\cr
&\Bigl(-\pdx \Pi(x)+
\U(x) -i\pi \rho(x) \Bigr)^{q+1} \biggr] \cr&=\int\sigma(y) y^{q}dy }.\eqn\sohr
$$
To prove our conjecture we have to show that the action $S[\rho, \sigma]$
on the trajectory
satisfying the boundary conditions \bcnd \ does have the property \sohr \
as a consequence of the equations of motion \euler.
To see that this is indeed the case, we introduce the set of time
dependent functions on the phase space (here $v(x) = \pdx \Pi(x)$)
$$\eqalign{H_{q}(\Pi, \rho, t)={1\over{2\pi i t (q+1)}}\int
dx\biggl\{&\Bigl(x-\bigl(v(x)-i\pi\rho(x)\bigr)t\Bigr)^{q+1} \cr
-&\Bigl(x-\bigl(v(x)+i\pi\rho(x)\bigr)t\Bigr)^{q+1}\biggr\}.} \eqn\hk$$
Using \euler, it is easy to check that
$${dH_{q}\over{dt}}=0$$
on the equations of motion of our original Hamiltonian \ham.
So we have found an infinite number of conservation laws corresponding to
the problem. Moreover,
$$\eqalign{H_{q}(\Pi, \rho, t=1)={1\over{2\pi i  (q+1)}}\int
dx\biggl\{&\Bigl(x-\bigl(v(x)-i\pi\rho(x)\bigr)\Bigr)^{q+1} \cr
-&\Bigl(x-\bigl(v(x)+i\pi\rho(x)\bigr)\Bigr)^{q+1}\biggr\},} $$
and
$$H_{q}(\Pi, \rho, t=0)=\int\sigma(y) y^{q}dy.$$
Now, since $H_{q}$ are the constants of motion, we conclude that \sohr \
does hold, and, therefore, \kaldef \ is proved.

\chapter{\bf Induced QCD With an Arbitrary Potential}

Induced QCD is the model defined by the partition
function \refmark{\kasmig}
$$\eqalign{
{\cal Z}=\int {\cal D}\Phi(x){\cal D}U_{\mu}(x)
\exp&\sum_{x}  N  {\rm Tr} \bigl[ -U\bigl(\Phi (x)\bigr) + \cr
&\sum_{\mu }  \Phi(x)
U_{\mu}(x)  \Phi(x+\mu a) U^{\dagger}_{\mu}(x) \bigr].}$$
The link fields $U_{\mu}(x)$ can be integrated out giving the effective
theory of the field $\Phi(x)$, with the action
$$ \eqalign{S=\!\!\  &\sum_{x,i}  \Bigl[-N U\bigl(\phi_i(x)\bigr)\Bigr]  +\!\!
\sum_{x,i\neq j} \ln |\phi_i(x)-\phi_j(x)| \cr + \!\!
&\sum_{x, \mu} \ln\Bigl[I\bigl(\phi(x), \phi(x+\mu a)\bigr)\Bigr]  \cr
} .  \eqn\effac
$$
In the large N limit the partition function of this effective theory is
dominated by a translationally invariant saddle point which
obeys the equation
$${1\over 2}U^{\prime}(\phi_{i})={1\over N}
\sum_{j\neq i}{1\over{\phi_{i}-\phi_{j}}}
+DN {\partial F_{0}(\phi, \chi)\over{\partial \phi_{i}}}\biggm|_{\phi=\chi}.
\eqn\sp$$
Using \kaldef \ and $N{\partial S \over
{\partial\phi_{i}}}=\bigl(\pdx {\delta S\over{\delta \rho(x)}}\bigr)\bigm|_
{x=\phi_{i}}=
\bigl(\pdx \Pi(x)\bigr)\bigm|_{x=\phi_{i}}=v(\phi_{i})$,
one can rewrite it in the form
$$v(\phi_{i})={1\over {2D}}U^{\prime}(\phi_{i}) - \phi_{i}+ {D-1\over D}
\sum_{j\neq i}{1\over{\phi_{i}-\phi_{j}}}
$$
or, equivalently,
$$ v(x) = {1\over {2D}}U^{\prime}(x) - x + {D-1\over{D}}{\cal P}\int{\rho(y)
 d y \over{x-y}}. \eqn\mfea $$
Migdal \refmark{\migd} has converted this into a nonlinear {\it integral}
equation for $\rho(x)$.
On the other hand, from the results of section 2 it follows that the problem
can be reduced to a {\it differential} equation. Indeed,
$$ v(x) = {\rm Re} f(x, t=0) = - {\rm Re} f(x, t=1)$$
where $f(x, t)$ is the solution of the boundary problem
$$\eqalign{
&{\partial f\over{\partial t}}+ f \pdx f =0 \cr
&\Im f(x, t=0)=\Im f(x, t=1)=\pi\rho(x) \cr} \eqn\mfeb $$
Equations \mfea \ and \mfeb \ provide a description of
induced QCD at large N.

Let us study the simplest case $D=1$. Then for a nonsingular potential $U(x)$
\footnote*{Let us, however, notice that for some singular potentials (say,
those considered in \refmark{\migphase})\ the analyticity hypothesis \parsol \
does not hold and one has to study the general equations \mfea \ and \mfeb.}
$$G_{+}(x)=x+f(x)=
x+v(x)+i\pi\rho(x)={1\over 2}U^{\prime}(x)+i\pi\rho(x), \eqn\kmd $$
and we obtain the
constraints determining the saddlepoint of one dimensional
induced QCD in the form
$$\cases {G_{+}(G_{-}(x))=G_{-}(G_{+}(x))=x \cr
        \Re G_{\pm}(x)={1\over 2}U^{\prime}(x) \cr
        \Im G_{\pm}(x)=\pm \pi \rho(x)}. \eqn\kmf$$
Note that in this case $G_{+}(x) = \overline{G_{-}(x)}$.
Moreover, since $D=1$ induced QCD is equivalent to the Hermitian
 matrix model
on a discretized line, one can use \kmf \  to investigate the properties of
this model.

In fact, it is easy
to construct a number of exact solutions to \kmf \ using the following device.
Consider an arbitrary real analytic symmetric function $P(z,w)=P(w,z)$
(say, a polynomial) of two variables, $z$ and
$w$. If we define $G_{+}$ and $G_{-}$ as the appropriate roots of
$$P(x, G_{+}(x))=P(G_{-}(x), x)=0,$$
then, by construction, $G_{+}(G_{-}(x))=x$, and
the roots $G_{+}$ , \  $G_{-}$ can be chosen
to be complex conjugate, \ie \ be the solutions of \kmf. To clarify
this point, let us consider an example
$$P(x, y)= x^2 +y^2 -{\rm m}^{2}x y +\mu.\eqn\quad$$
For this $P(x,y)$
$$G_{\pm}(x)={{\rm m}^{2}\over 2}x \pm i \sqrt{\mu - {\mu^{2}x^{2}\over 4}}$$
so that $\pi \rho(x)= \sqrt{\mu - {\mu^{2}x^{2}\over 4}}$ \  (where $\mu$
is fixed by $\int \rho(x) dx =1$ to be $\mu=\sqrt{{\rm m}^{4}-4}$)
-the semicircular law, which is known to solve induced QCD with the
quadratic potential $U(x)={\rm m}^{2}x^{2}/2 \,\, $  \refmark{\gross}
. One can verify
explicitly that \kmf \ indeed holds for these $G_{\pm}$.

To obtain a more general solution,
 we can perturb \quad \ by monomials of higher order.
Then
the relation $G_{+}(G_{-}(x))=G_{-}(G_{+}(x))=x$ will not be violated, at least
for small enough coefficients in front of these monomials.
The simplest generalization is
$$P(x,y)=x^2 +y^2 -b x^{2}y^{2} -{\rm m}^{2}x y +\xi(x+y) +\alpha.$$
The corresponding potential and density are given by
$$\eqalign{ U^{\prime}(x)&={{\rm m}^{2}x - \xi \over {1- b x^{2}}} \cr
\pi \rho(x)&={1\over {1- b x^{2}}}\sqrt{(1- b x^{2})(x^{2} +\xi x+\alpha)-
{1\over 4}({\rm m}^{2}x - \xi)^{2}} \cr}\,\,.    \eqn\model $$
To see the physical meaning of this solution, let us consider its continuum
limit. In the continuum limit we must scale
$$  \Phi \to {\varphi(t) \over \sqrt{a}}; \,\, \,\, m^2 = 2 + a^2 M^2 ;
b=B a^3;\,\,$$ $$ \,\, \xi=\Xi a^{3/2};
\,\, \,\, U(\Phi)= \Phi^{2}+a W( \varphi) \,\,,
$$
where $a$ is the lattice spacing.
As $a \rightarrow 0$,
the partition function of induced QCD reduces to
$$Z= \int {\cal D} \varphi(t)
e^{ - N\int dt\bigl[    {\rm Tr} \{  {M^2\over 2}\varphi(t)^2
+ W(\varphi(t)) +\oh (\partial_t \varphi(t))^2 \}\bigr] }  \,\, . \eqn\rr
$$
In our example
$$W^{\prime}(\varphi)={M^2 \varphi +2 B \varphi^{3} -\Xi\over
{1-B a^2 \varphi^{2}}}.$$
We see that as $a\rightarrow0$,
$$W(\varphi)\rightarrow {M^2 \varphi^{2}\over 2}+ {B \varphi^{4}\over 2}-\Xi
\varphi
$$ which is a general quartic potential.
Note that our solution applies not only in the continuum limit, but for
finite lattice spacings as well.

Therefore, it is possible to find nontrivial potentials, which have a
well-defined continuum limit and, at the same time, admit an exact solution
of the model.

\chapter{\bf Conclusions}

We have shown that the large N behaviour of the Itzykson -- Zuber
integral, which,
by itself, is not dominated by any of its $N!$ saddlepoints on
the group manifold $U(N)$, is nevertheless controled by a well defined
classical trajectory in the space of eigenvalues. We obtained the
representation of its asymptotics in terms of that trajectory and
reduced the problem to a set of two algebraic constraints. It turned out
that the analysis of induced QCD is the problem, dual to
the computation of the Itzykson -- Zuber integral:  to solve induced QCD
we have to find $\rho(x)$, given $U(x)$ in \kmf, while in
order to compute the integral we must solve for $U(x)$, when $\rho(x)$ is
known. We demonstrated how some particular solutions of this problem
can be obtained.

Finally, it is necessary to make the following remark. Makeenko
\refmark{\makeenko} \ claimed that induced QCD in $D$ dimensions
with an even potential $U(x)$ has the same saddle point as the one-matrix
model with the potential ${\tilde U}(x)$ related to $U(x)$ by
$${\tilde U}^{\prime}(x)={1\over{2D-1}}\Bigl((D-1)U^{\prime}(x)+
D\sqrt{\bigl(U^{\prime}(x)\bigr)^{2} + 4(1-2D)x^{2}}\Bigr).
\eqn\makmod $$
However, it appears that there are some counterexamples to this statement.
Let us, say, consider the $D=1$ case. Choose
$$U(x)= {{\rm m}^{2} x^{2} \over {2}} + { g x^{4}\over {4}}.$$
It is easy to find the solution of this model up to the first order in $g$.
One can do it using \model \ with $\xi=0$ and $b={g\over {{\rm m}^{2}}}\ll 1.$
Then the eigenvalue density $\rho(x)$ reduces to
$$ \rho(x)={1\over{\pi}}(cx^{2}+d)\sqrt{a-x^{2}},$$
where
$$c={g\over{2\mu{{\rm m}^{2}}}}({{\rm m}^{4}}-2) +{\cal O}(g^{2}),
\,\,\  d= {\mu \over{2}}+ g {{{\rm m}^{2}} \over {\mu^{2}}} +{\cal O}(g^{2})
\eqn\pertone $$ and
$\mu=\sqrt{{{\rm m}^{4}}-4}$. Then the constant $a$ is fixed by the
requirement $\int\rho(x)dx=1$. It can be checked that {\it the same
result follows independently from the perturbation theory
for Migdal's equation} \refmark{\migd}. However, the results following
from \makmod \ do not agree with that. Indeed, the corresponding one-matrix
potential would be
$$ {\tilde U}(x)= {\mu x^{2}\over{2}}+ {g {{\rm m}^{2}}\over {4\mu}}x^{4}
+{\cal O}(g^{2}).$$
This matrix model is easy to solve exactly \refmark{\bipz}.
Expanding the result in $g$ one gets
$$\rho(x)={1\over{\pi}}({\tilde c}x^{2}+{\tilde d})\sqrt{{\tilde a}-x^{2}},$$
with
$${\tilde c}= {g {{\rm m}^{2}} \over {2\mu}}+{\cal O}(g^{2}), \,\,\
{\tilde d}= {\mu \over{2}}+ g {{{\rm m}^{2}} \over {\mu^{2}}} +
{\cal O}(g^{2}).$$
We see that ${\tilde c}\ne c$ already in the first order of
perturbation theory, indicating problems with \makmod.

\chapter{\bf Acknowledgements}

I am most indebted to David Gross for suggesting this problem, as well
as for his advice, support and interest in this work. I am grateful to S.
Shatashvili for the idea of the proof in section 3, and to A.A.
Migdal for discussions.

\refout

\end